\documentclass[12pt]{article}

\input epsf
 \usepackage{amssymb,wrapfig}
\usepackage{xypic,multirow}
\usepackage[matrix,arrow,curve]{xy}

\begin{document}
\addtolength{\baselineskip}{.5mm}
\newlength{\extraspace}
\setlength{\extraspace}{1.5mm}
\newlength{\extraspaces}
\setlength{\extraspaces}{2mm}

\def\nn{{\cal N}}
\def\rr {{\Bbb R}}
\def\cc {{\Bbb C}}
\def\pp {{\Bbb P}}
\def\zz {{\Bbb Z}}
\def\del {\partial}
\def\cy {Calabi--Yau}
\def\ka {K\"ahler}
\newcommand{\inv}[1]{{#1}^{-1}} 

\newcommand{\Section}[1]{\section{#1} \setcounter{equation}{0}}

\makeatletter
\@addtoreset{equation}{section}
\makeatother
\renewcommand{\theequation}{\thesection.\arabic{equation}}

\begin{center}

{\hfill SLAC--PUB--11511\\ \hfill SU--ITP--05/29}

{\Large \bf{Complex/Symplectic Mirrors}}

\bigskip

Wu--yen Chuang$^{1,2}$, Shamit Kachru$^{1,2}$ and Alessandro Tomasiello$^1$

\medskip 

$^1$ {\it ITP, Stanford University, Stanford, CA 94305, USA}

$^2$ {\it SLAC, Stanford University, Menlo Park, CA 94309, USA }

\bigskip
{\bf Abstract}
\end{center}
We construct a class of symplectic non--K\"ahler and complex non--K\"ahler
string theory vacua, extending and providing
evidence for an earlier suggestion by Polchinski and Strominger.
The class admits a mirror pairing by construction. Comparing hints from
a variety of sources, including ten--dimensional supergravity and KK reduction
on SU(3)--structure manifolds, suggests a picture in which
string theory extends Reid's fantasy to connect classes of both complex
non-K\"ahler and symplectic
non-K\"ahler manifolds.


\section{Complex and symplectic vacua}
The study of string theory on
\cy\ manifolds has provided both the most popular
vacua of the theory, and some of the best tests
of theoretical ideas about its
dynamics.
Most manifolds, of course, are not \cy. What 
is the next simplest class for theorists to explore? 

The answer, obviously, depends on what the definition of ``simplest'' is.
However, many leads seem to be pointing to the same suspects. First
of all, it has been
suggested long ago \cite{ps}
that type II vacua exist, preserving $\nn=2$
supersymmetry ({\it the same} as for \cy's), on manifolds which are complex
and non--\ka\ (and enjoy vanishing $c_1$). \cy\ manifolds are simultaneously complex and symplectic,
and mirror symmetry can be viewed as an exchange of these two
properties \cite{kontsevich}.
The same logic seems to suggest that the proposal of \cite{ps} should
also include symplectic non--\ka\ manifolds
as mirrors of the complex non--\ka\ ones. Attempts
at providing mirrors of this type (without using a physical interpretation)
have indeed already been made in \cite{sty,st}.

In a different direction, complex non--\ka\ manifolds have also featured in 
supersym\-metry--preserving vacua of supergravity, already in 
\cite{andy}. More recently the general conditions
for preserving $\nn=1$ supersymmetry in supergravity have been reduced to 
geometrical conditions
\cite{gmpt2}; in particular, the manifold has to be generalized complex
\cite{hitchin}. The most prominent examples of generalized complex manifolds
are complex and symplectic manifolds, neither necessarily \ka. It should
also be noted that complex and symplectic manifolds seem to be natural
in topological strings.

In this paper we tie these ideas together. We find that vacua of the type
described in \cite{ps} can be found for a large class of complex non--\ka\
manifolds in type IIB and symplectic non--\ka\ manifolds in type
IIA, and observe
that these vacua come in mirror pairs. Although these vacua are not
fully amenable to ten--dimensional supergravity analysis for reasons
that we will explain
(this
despite the fact that they preserve $\nn=2$ rather than $\nn=1$ supersymmetry), this is in agreement with
the supergravity picture that all (RR) SU(3)--structure IIA vacua are
symplectic \cite{gmpt}, and all IIB vacua are complex \cite{dall,frey,gmpt}, 
possibly suggesting a deeper structure.

In section \ref{sec:vacua}, in an analysis formally identical to \cite{ps},
we argue for the existence of the new vacua.
In section \ref{sec:geom} we show that the corresponding internal manifolds are not \cy\  but rather complex or symplectic. More specifically, in both
theories, they are obtained from a transition that does not preserve
the \cy\ property. As evidence for this, we show that
the expected physical spectrum agrees with the one obtained on
the proposed manifolds. The part of this check that concerns
the massless spectrum is straightforward;
we can extend it to low-lying massive fields by combining results from
geometry \cite{dish} and KK reduction on manifolds of SU(3) structure.
We actually try in section \ref{sec:spec}
to infer from our class of examples a few properties which
should give more control over this kind of KK reduction. Specifically, we
suggest that the lightest massive fields should
be in correspondence with pseudo--holomorphic
curves or pseudo--Special--Lagrangian three--cycles (a notion we will
define at the appropriate juncture).

Among the motivations for this paper were also a number of more grandiose
questions about the effective potential of string theory. One
of the motivations
for mathematicians to study the generalized type of transition we consider
in this paper is the hope that many moduli spaces actually happen
to be submanifolds of a bigger moduli space, not unlike \cite{reid}
the realization
of the various 19--dimensional
moduli spaces of algebraic K3's as submanifolds
of the 20--dimensional moduli space of abstract K3's. It might be that
string theory provides a natural candidate for such a space, at least for the
${\cal N}=2$ theories,
whose points would be all
SU(3)--structure manifolds (not necessarily complex or symplectic),
very possibly augmented by non--geometrical points \cite{Wati}. We would
not call it a moduli space, but rather a configuration space:
on it, a potential would be defined, whose zero
locus would then be the moduli space of ${\cal N}=2$ supersymmetric
string theory vacua,
including in particular the complex and symplectic
vacua described here. In this context, what this paper is studying is a
small neighborhood where the moduli space of ${\cal N}=2$
non-K\"ahler compactifications
meets up with the moduli space of
Calabi-Yau compactifications
with RR flux,
inside this bigger configuration space of manifolds.

\section{Four--dimensional description of the vacua}
\label{sec:vacua}

We will now adapt the ideas from \cite{ps} to our needs.
The strategy is as follows. We begin by compactifying the IIB and IIA
strings on Calabi-Yau threefolds,
and we switch on internal RR fluxes, $F_3$ in IIB and
$F_4$ in IIA (our eventual interest will be the case where the theories
are compactified on mirror manifolds ${\cal M}$ and ${\cal W}$, 
and the fluxes are mirror
to one another). As also first noted in \cite{ps}, this will make the
four--dimensional $\nn=2$ supergravity gauged; in particular, it will
create a potential on the moduli space. This potential has
supersymmetric vacua only at points where the \cy\ is singular. However, on
those loci of the moduli space new massless brane hypermultiplets have to
be taken into account, which will then produce the new vacua.

\subsection{The singularities we consider}
\label{sec:sing}
Let us first be more precise about the types of singularities we will consider.
In IIB, as we will review shortly, if we switch on
$F_3$ with a non--zero integral along a cycle $B_3$ of a Calabi-Yau ${\cal M}$,
a supersymmetric vacuum will exist on a point in moduli space
in which only the cycle $A_3$ conjugate to $B_3$ under intersection pairing
shrinks. It is often the case that several cycles shrink
simultaneously, with effects that we will review in the next section,
but there are definitely examples in which a single $B$ cycle shrinks.
These are the cases we will be interested in. (We will briefly
explain in section \ref{sec:ncytrans} how this condition could be 
relaxed.)

In IIA, switching on $F_4$ with a non--zero integral on a four--cycle
$\tilde A_4$ of ${\cal W}$ will 
generate a potential which will be zero only in points in
which the quantum--corrected volume of the conjugate two--cycle $\tilde B_2$
(the Poincar\'e dual to $F_4$) vanishes. This will happen on a wall
between two birationally equivalent \cy's, connected by a flop of
$\tilde B_2$.
These points will be mirror to the ones we described above for IIB.

The converse is not always true: there can be shrinking three--cycles which are
mirror to points in the IIA moduli space in which the quantum volume
of the whole \cy\ goes to zero. These walls separate geometrical and
Landau--Ginzburg, or, hybrid, phases. One would obtain a vacuum at
such a point by switching on
$F_0$ instead of $F_4$, for instance. The example
discussed in \cite{ps} (the quintic) is precisely such a case.
Since in the end we want to give
geometrical interpretations to the vacua we will obtain, we will restrict
our attention only to cases in which a curve shrinks in ${\cal W}$ -- 
that is, when a flop happens. Although this is not
strictly necessary for IIB, keeping mirror symmetry in mind we will
restrict our attention to cases in which the stricter IIA condition is valid,
not only the IIB one: in the mirror pairs of interest to us, 
the conifold singularity in 
${\cal M}$ is mirror to a flop in ${\cal W}$. It would be interesting, of 
course, to find the IIA mirrors to all the other complex non--\ka\ manifolds
in IIB. 

Looking for flops is not too difficult, as there is
a general strategy. If the \cy\ ${\cal W}$ is realized as hypersurface in
a toric manifold $V$,
the ``enlarged \ka\ moduli space'' \cite{greene,agm}
(or at least, the part
of it which comes from pull--back of moduli of $V$) is a toric
manifold $W_K$ itself. The cones of the fan of $W_K$
are described by different triangulations of the cone over the
toric polyhedron of $V$. Each of these cones will be a phase
\cite{Witten}; there will be
many non--geometrical phases (Landau--Ginzburg or hybrid). Fortunately,
the geometrical ones are characterized as the triangulations of the toric
polyhedron of $V$ itself (as opposed to triangulations of the cone over it).
This subset of cones gives an open set in $W_K$ which is called the
``partially enlarged'' \ka\ moduli space.
This is not the end of the story, however. In many examples, it will happen
that a flop between two geometrical phases will involve more than one curve
at a time, an effect due to restriction from $V$ to ${\cal W}$.
Worse still, these curves might have relations, and sometimes there is
no quick way to determine this.
Even so, we expect that there should be many cases in which a single
curve shrinks (or many, but without relations).

Such an example is readily found in the literature \cite{mv,lsty}:
taking ${\cal W}$ to be an elliptic fibration
over ${\Bbb F}_1$ (a \cy\ whose Hodge numbers are $h^{1,1}=3$ and
$h^{2,1}=243$), there is a point in moduli space in which a single curve
shrinks (see Appendix \ref{app:toric} for more details). By counting of 
multiplets and mirror symmetry, on the mirror
${\cal M}$ there will be a single three--cycle which will shrink. This implies
that the mirror singularity will be a conifold singularity. Indeed,
it is a hypersurface singularity, and as such the shrinking cycle
is classified by the so--called Milnor number. This
has to be one if there is a single shrinking cycle, and the only
hypersurface singularity with Milnor number one is the conifold.

\subsection{Gauged supergravity analysis}
\label{sec:gauged}
  After these generalities, we will now show how turning on fluxes drives
the theory to a conifold point in the moduli space; more importantly,
we will then show how including the new massless hypermultiplets generates
new vacua. We will do this in detail in the IIB theory on ${\cal M}$, 
as its IIA counterpart
is then straightforward. The analysis is formally identical to the one in
\cite{ps} (see also \cite{michelson,Ferrara}); the
differences have been explained in the previous subsection.

As usual, define the symplectic basis of three--cycles $A^I$, $B_J$  and
their Poincar\'e duals $\alpha_I$, $\beta^I$ such that
\begin{equation}
A^I \cdot B_J = \delta^I{}_J\ , \qquad
\int_{A^J} \alpha_I = \int_{B_I} \beta^{J} = \delta_I{}^J
\end{equation}
along with the periods $X^I = \int_{A^I} \Omega$ and $F_I = \int_{B_I}\Omega$.
Additionally, the basis is taken so that the cycle of interest described
in subsection \ref{sec:sing} is $A=A^1$.

When $X^1=0$, the cycle $A^1$ degenerates to the zero
size and ${\cal M}$ develops a conifold singularity. 
By the monodromy argument,
the symplectic basis $(X^1, F_1)$ will transform as follows when
we circle the discriminant locus in the complex moduli space
defined by $X^1=0$:
\begin{equation}
X^1 \to X^1 \ \ \ \  F_1 \to F_1 + X^1\ .
\end{equation}
From this we know $F_1$ near the singularity:
\begin{equation}
F_1 = {\rm constant} + \frac{1}{2 \pi i} X^1 {\rm ln} X^1 + \dots
\end{equation}
The metric on the moduli space can be calculated from the formulae
\begin{equation}
\mathcal{G}_{I \bar{J}} = \partial_I \partial_{\bar{J}} K_V\ , \qquad
K_V= - \ln i ( \bar{X}^I F_I - X_I \bar{F}^I)\ .
\end{equation}
Therefore we obtain
\begin{equation}
\mathcal{G}_{1 \bar{1}} \sim \ln( X^1 \bar{X^1})\ .
\end{equation}

Now, the internal flux we want to switch on is $F_3=n_1 \beta^1$. The
vectors come from
\begin{equation}
F_5=F_2^I\wedge \alpha_I - G_{2,I} \wedge \beta^I\ ,
\end{equation}
where the $F_2^{I}$ ($G_{2I}$) is the electric (magnetic) field
strength. The Chern-Simons coupling in the IIB supergravity action is then
\begin{equation}
\epsilon^{ij} \int_{M_4\times CY} \tilde{F}_5 \wedge H_3^{i} \wedge B_2^{j}=
n_1\int_{M_4} F_2^1 \wedge B_2
\end{equation}
where $M_4$ is the spacetime.
By integration by parts, and since $B_2$ dualizes to one of
the (pseudo)scalars in the universal hypermultiplets, we see that the
latter is gauged under the field $A^1$ whose field strength is $dA^1=F_2^1$.

The potential is now given by the ``electric'' formula
\begin{equation}
  \label{eq:elec}
V= h_{uv} k_I^{u} k_J^{v} \bar{X}^{I} X^{J} e^{K_V} + (U^{IJ} - 3
\bar{X}^{I} X^{J} e^{K_V} ) \mathcal{P}_I^{\alpha} \mathcal{P}_J^{\alpha}\
\end{equation}
where
\begin{equation}
U^{IJ}= D_a X^I g^{a\bar b} D_{\bar b} X^J \
\end{equation}
and the ${\cal P}^\alpha$ are together the so--called Killing prepotential,
or hypermomentum map.
In our situation only the flux over $B_1$ is turned on, and the Killing
prepotential is given by
\begin{equation}
\mathcal{P}_1^{1}=\mathcal{P}_1^{2}=0\ ; \quad
\mathcal{P}_1^{3}= -e^{\tilde{K}_H} n_{eI}^{(2)} = - e^{2\phi} n_{eI}^{(2)} 
\end{equation}
where $\phi$ is the dilaton.
The potential will
then only depend
on the period of the dual $A^1$ cycle, call it $X^1$:
\begin{equation}
V\sim \frac{(n_1)^2}{\ln X^1 \bar X^1}\ .
\end{equation}
The theory will thus be driven to the conifold point where $X^1=0$.


This is not the end of the story: at the singular point, one has a new massless
 hypermultiplet $B$ coming from a brane wrapping the shrinking cycle $A^1$.
 The world--volume coupling between the
D3-brane and $F_5$ gives then $\int_{\rr\times A^1} A_4 = \int_{\rr} A^1$,
where $\rr$ is the worldline of the resulting light particle in $M_4$. (The
coincidence between the notation for the cycle $A^1$ and the corresponding
vector potential $A^1$ is rather unfortunate, if standard.)

This means that both the universal and the brane hypermultiplet are
charged under the same vector; we can then say that they are all
electrically charged and still use the electric formula for the potential
(\ref{eq:elec}), with the only change being that the Killing prepotential is
modified to be
\begin{equation}
  \label{eq:kprep}
\mathcal{P}_1^\alpha
= \mathcal{P}_1^\alpha |_{B=0} + B^{+} \sigma^\alpha B\ ;
\end{equation}
the black hole hypermultiplet is an $SU(2)$ doublet with components
$(B_1, B_2)$.
Loci on which the ${\cal P}^\alpha$'s are
zero are new vacua: it is easy to see that they are given by
\begin{equation}
  \label{eq:vacua}
B=( (e^{\tilde{K}_H} n_1)^{1/2}, 0) =(e^{\phi} n_1^{1/2} ,0)
\ .
\end{equation}

The situation here is similar to \cite{ps}: the expectation value
of the new brane hypermultiplet is of the order $g_s= e^{\phi}$.
So, as in that paper, the two requirements that $g_s$ is small and
that $B$ be small (the expression for the ${\cal P}^\alpha$ is a
Taylor expansion and will be modified for large $B$) coincide, and
with these choices we can trust these vacua. After the Higgsing
the flat direction of the potential, namely, the massless
hypermultiplet $\tilde{B}_0$, would be a linear combination of the
brane hypermultiplet and the universal hypermultiplet while the
other combination would become a massive one $\tilde{B}_m$.

\subsection{The field theory capturing the transition}
\label{sec:hyper}

It is useful to understand the physics of the transition from a 4d
field theory perspective, in a region very close to the transition point
on moduli space.  
While this analysis is in principle a simple limit of the gauged supergravity
in the previous subsection, going through it will 
both provide more intuition and also allow us to infer some additional
lessons.
In fact, in the IIB theory with $n_1$ units of RR
flux, the theory close to the transition point (focusing on the
relevant degrees of freedom) is simply a U(1) gauge
theory with two charged hypers, of charges $1$ and $n_1$.

Let us focus on the case $n_1=1$ for concreteness.  
Let us call the ${\cal N}=1$ chiral multiplets in the two hypers
$B, \tilde B$ and $C, \tilde C$.  
In ${\cal N}=1$ language, this theory has a superpotential
\begin{equation}
W \sim \tilde B \varphi B + \tilde C \varphi C
\end{equation}
where $\varphi$ is the neutral chiral multiplet in the ${\cal N}=2$ 
U(1) vector multiplet.
It also has a D-term potential
\begin{equation}
\vert D \vert^2 \sim (|\tilde B|^2 - |B|^2 + |\tilde C|^2 - |C|^2)^2~.
\end{equation}
There are two branches of the moduli space of vacua: a Coulomb branch
where $\langle \varphi \rangle \neq 0$ and the charged matter fields
vanish, and a Higgs branch where
$\langle \varphi \rangle = 0$ and the hypers have non-vanishing vevs
(consistent with $F$ and $D$ flatness).
The first branch has complex dimension one, the second has quaternionic
dimension one.  These branches meet at the point where all fields
have vanishing expectation value.

At this point, the theory has an SU(2) global flavor symmetry.  This
implies that locally, the hypermultiplet moduli space
will take the form $\cc^2/\zz_2$ \cite{Seiberg}.  In fact, the precise
geometry of the hypermultiplet
moduli space, including quantum corrections, 
can then be determined by a variety of arguments 
\cite{Seiberg,SS} (another type of argument \cite{Vafa} implies the
same result for the case where the hypermultiplets coming from 
shrinking three--cycles in IIB).
The result is the following.  Locally, the quaternionic space reduces
to a hyperK\"ahler manifold which is an elliptic fibration,
with fiber coordinates $t,x$ and a (complex) base coordinate $z$.  Let
us denote the K\"ahler class of the elliptic fiber by $\lambda^2$.  
Then, the metric takes the form
\begin{equation}
\label{metis}
ds^2 = \lambda^2 \left( V^{-1} (dt - {\bf A} \cdot {\bf dy})^2 + V ({\bf dy})^2
\right)
\end{equation}
where ${\bf y}$ is the three-vector with components $(x,{z \over \lambda}, 
{{\bar z}\over \lambda})$. 
Here, the function $V$ and the vector of functions ${\bf A}$ are given by
\begin{equation}
\label{vis}
V = {1\over 2\pi} \sum_{n=-\infty}^{\infty}\left( {1\over {\sqrt{(x-n)^2 
+ {|z|^2 \over \lambda^2}}}} - {1\over |n|} \right) ~+~{\rm constant} 
\end{equation}
and
\begin{equation}
\label{ais}
{\bf \nabla} \times {\bf A} ~=~{\bf \nabla} V~. 
\end{equation}
This provides us with detailed knowledge of the metric on the 
hypermultiplet moduli space emanating from the singularity, though
it is hard to explicitly map the flat direction to a combination
of the universal hypermultiplet and the geometrical parameters of 
${\cal M}'$ or ${\cal W}'$.  We shall discuss some qualitative aspects
of this map in \S3.3.
For the reader who is confused by the existence of a Coulomb branch at all,
given that e.g. in the IIB picture $F_3 \neq 0$, we note that the
Coulomb branch will clearly exist on a locus where $g_s \to 0$ (since
the hypermultiplet vevs must vanish).  This is consistent with 
supergravity intuition, since in the 4d Einstein frame, the energetic
cost of the RR fluxes vanishes as $g_s \to 0$.

\section{Geometry of the vacua}
\label{sec:geom}

We will first of all show that the vacua obtained in the previous section
cannot come from a transition to another \cy. To this aim, in the next
subsection we will review \cy\ extremal transitions. We will then proceed
in subsection \ref{sec:ncytrans}
to review the less well--known {\it non}--\cy\ extremal transitions, and
then compare them to the vacua we previously found 
in subsection \ref{sec:compare}.

\subsection{\cy\ extremal transitions}
\label{sec:cytrans}

\cy\ extremal transitions sew together moduli spaces for Calabi--Yaus
whose Hodge numbers differ; let us quickly review how. For more details
on this physically well--studied case, the reader might want to consult
\cite{cgh,gms,cggk,greene}.

 Consider IIB theory on a \cy\ ${\cal M}$. (Some of the explanations in this
paper are given in the IIB case only, whenever the IIA case would be
an obvious enough modification). 
Suppose that at
a particular point in moduli space, ${\cal M}$ develops $N$ nodes
(conifold points)
by shrinking as many three--cycles $A_a$, $a=1,\ldots,N$, and that these
three--cycles satisfy $R$ relations
\begin{equation}
\sum_{a=1}^N r_i^a A_a ~=~0,~~i=1,\cdots,R
\end{equation}
in $H_3$.
We are {\it not} using the same notation for the index on the cycles
as in section \ref{sec:vacua},
as these $A_a$ are not all elements of a
basis (as they are linearly dependent).
Notice that it is already evident that this case is precisely the one
we excluded with the specifications in section \ref{sec:sing}.
To give a classic example \cite{cgh},  there is a known
transition where ${\cal M}$ is the quintic,
$N=16$ and $R=1$.
Physically, there will be $N$ brane hypermultiplets $B_a$
becoming massless
at this point in moduli space. Vectors come from $h^{2,1}$; since the
$B_a$ only span $N - R$ directions in $H^3$, they will be charged under
$N-R$ vectors $X^A$ only, $A=1,\ldots N-R$. Call the matrix of
charges $Q_A^a$, $A=1,\ldots,N-R$, $a=1,\ldots N$.

In this case, when looking for vacua, we will still be
setting the Killing prepotential ${\cal P}_a$ (which is a simple
extension of the one in (\ref{eq:kprep})) to zero: 
the flux is now absent, and the
$B^2$ term now reads
\begin{equation}
{\cal P}_A=\sum_a Q^a_A B_a^+ \sigma^\alpha B_a\ .
\end{equation}
Notice that we have switched no flux on in this case; crucially, ${\cal P}=0$
now will have an $R$--dimensional space of solutions, due to the relations.

Let us suppose this new branch is actually the moduli space for a new \cy.
This new manifold would have
$h^{2,1}-(N-R)$ vectors, because all the $X^A$ have been Higgsed; and
$h^{1,1}+R$ hypers, because of the $N$ $B_a$, only $R$ flat directions
have survived.

This is exactly the same result one would get from a small
resolution of all the $N$ nodes. Indeed, let us call the \cy\ resulting
from such a procedure ${\cal M}'$, and let us compute its Betti numbers.
It is actually simpler to first consider a case in which
a single three--cycle undergoes surgery\footnote{This is a purely topological 
computation; in a topological context,
an extremal transition is called a surgery, and we will use this term
when we want to emphasize we are considering purely the topology of the
manifolds involved.}, which is the case without relations specified in
section \ref{sec:sing}; we will go back to the \cy\ case, in which relations
are necessary, momentarily.

\begin{figure}[ht]
  \centering
  \begin{picture}(200,200)
\put(208,85){\small $C$}\put(-80,0){\epsfxsize=5in\epsfbox{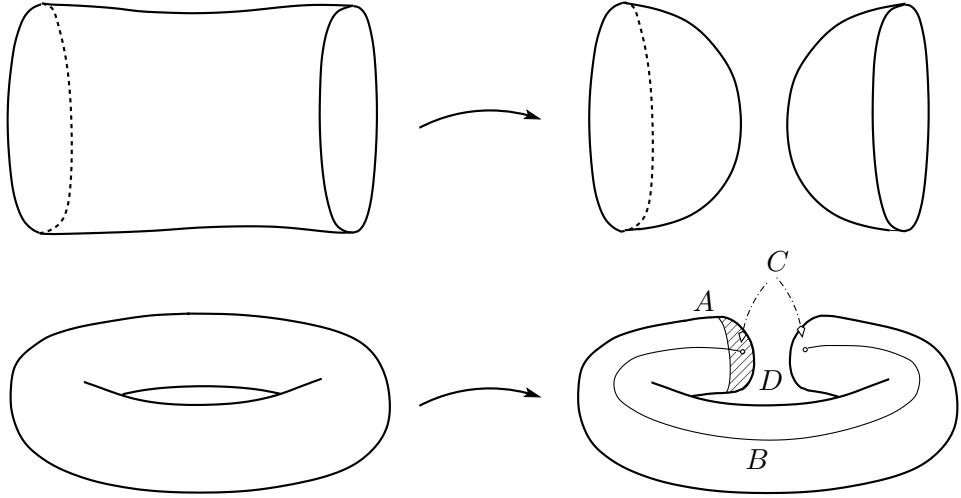}}
\put(200,10){\small $B$}\put(180,70){\small $A$}
\put(205,40){\small $D$}
  \end{picture}
  \caption{Difference between compact and non--compact surgery: in the
noncompact case (up), one loses an element in $H^1$
and one gains an element in $H^0$ (a connected component).
In the compact case (down), one loses an element in $H^1$ again, but the
would--be new element in $H^0$ is actually trivial, so $H^0$ remains the
same. This figure is meant to help intuition about the conifold
transition in dimension 6, where
$H^0$ and $H^1$ are replaced by $H^2$ and $H^3$. We also have depicted
various chains on the result of the compact transition, for later use.}
  \label{fig:trans}
\end{figure}

The result of this single surgery along a three--cycle is that
$H^3\to H^3-2$, $H^2\to H^2$.  This might be a bit surprising: one is used
to think that an extremal transition replaces
a three--cycle by a two--cycle. But this intuition comes from the noncompact
case, in which indeed it holds. In the compact case, when we perform a
surgery along a three--cycle, we really are also losing its conjugate
under Poincar\'e pairing; and we gain no two--cycle.
The difference is
illustrated in a low--dimensional analogue in figure
\ref{fig:trans}, in which $H^2$ and $H^3$
are replaced by $H^0$ and $H^1$.

Coming back to the \cy\ case of interest in this subsection, let us now
consider  $N$ shrinking three--cycles with $R$ relations.
First of all $H^3$ only changes by $2(N-R)$, because this is the number
of independent cycles we are losing. But this is not the only effect on the
homology. A relation can be viewed as
a four--chain $F$ whose boundary is $\sum A_a$. After surgery, the boundary
of $F$ by definition shrinks to points; hence $F$ becomes a four--cycle
in its own right. This gives $R$ new elements in $H_4$ (or equivalently,
in $H^2$). The change
in homology is summarized in Table \ref{table}, along with the IIA case and,
more importantly, in a more general context that we will explain. By comparing
with the physical counting above, we find evidence that the new branches
of the moduli space correspond to new \cy\ manifolds obtained by extremal
transitions.

To summarize, \cy\ extremal transitions are possible without fluxes, but
they require relations among the shrinking cycles. This is to be
contrasted with the vacua in the previous section, where there are no
relations among the shrinking cycles to provide flat directions.  Instead,
the
flux (and resulting gauging) lifts the old Calabi-Yau moduli space
(as long as $g_s \neq 0$),
but makes up for this by producing a new branch
of moduli space (emanating from the conifold point or its mirror).

\subsection{Non--\cy\ extremal transitions}
\label{sec:ncytrans}

In this section we will waive the \cy\ condition to reproduce the vacua
of the previous section. This is, remember, a case in which cycles shrink
without relations. However, we will start with a review of results in the
more general case, to put in perspective both the case we will eventually
consider and the usual \cy\ case.

We will consider both usual conifold transitions, in which three--cycles
are shrunk and replaced by curves, and so--called reverse conifold transitions, in which the converse happens.\footnote{Implicit in the use of the word
``conifold'' is the assumption that several
cycles do not collapse together in a single point of the manifold ${\cal M}$.
More general cases are also interesting to consider, see for example
\cite{cggk}
for the complex case and \cite{st} for the symplectic case.}
As a hopefully useful shorthand, we will call
the first type a $3\to2$ transition and the second a $2\to3$.
Though the manifolds will no longer be (necessarily) \cy, we will still
call the initial and final manifold ${\cal M}$ and ${\cal M}'$ in the
$3\to2$ case (which is relevant for our IIB picture), and ${\cal W}$
and ${\cal W}'$ in the $2\to3$ case (which is relevant for our IIA picture).

We will first ask whether a $3\to2$ transition takes a complex, or symplectic,
${\cal M}$ into a complex, or symplectic, ${\cal M}'$, and then turn to
the same questions about ${\cal W}, ~{\cal W}'$ 
for $2\to 3$ transitions. These questions have to be phrased a bit more
precisely, and we will do so case by case.

It is also useful to recall at this point the definitions of symplectic
and complex manifolds, which we will do by embedding them in a bigger 
framework.
In both cases, we can start with a weaker concept 
called {\it $G$--structure}. By
this we mean the possibility of taking the transition functions on the 
tangent bundle of ${\cal M}$ to be in a group $G$. This is typically 
accomplished
by finding a geometrical object (a tensor, or a spinor) whose stabilizer
is precisely $G$. If we find a two--form $J$ such that $J\wedge J \wedge J$
is nowhere zero, 
it gives an Sp$(6,\rr)$ structure. In presence of a tensor $(1,1)$ tensor (one 
index up and one down) $j$ such that $j^2=-1$ (an almost complex 
structure), we speak of a Gl$(3,\cc)$ 
structure. For us the presence of both will be 
important; but we also impose a compatibility condition, which says that the
tensor $j_m{}^p J_{pn}$ is symmetric and of positive signature. This tensor
is then nothing but a Riemannian metric. The triple is 
an almost hermitian metric: this 
gives a structure Sp$(6,\rr)\cap $ Gl$(3,\rr)=$U(3). 

By themselves, these reductions of structure do not give much of a restriction 
on the manifold. 
But in all these cases we can now consider an appropriate integrability 
condition, a differential equation which makes the manifold with the
given structure more rigid. 
In the case of $J$, we can impose that $dJ=0$. In this case we say that the 
manifold is symplectic. For $j$, a more complicated condition (that we will 
detail later, when considering SU(3) structures) leads to complex manifolds.

Let us now consider a complex manifold ${\cal M}$ (which we will also take
to have trivial canonical class $K=0$). First order complex deformations are 
parameterized by
$H^1({\cal M},T)=H^{2,1}$. Suppose that for some value
of the complex moduli $N$ three--cycles shrink. Replace now these
$N$ nodes by small resolutions. The definition of small resolution, just like
the one of blowup, can be given locally around the node and then patched
without any problem with the rest of the manifold. So the new manifold
${\cal M}'$ is still complex.
Also, the canonical class $K$ is not modified by the transition because
a small resolution does not
create a new divisor, only a new curve.\footnote{
The conditions for $\nn=1$--preserving vacua in ten--dimensional type II
supergravity actually only require $c_1=0$. The role of this condition is
 less clear for example in the topological
string: for the A model it would seem to unnecessary, as there is no anomaly
to cancel; for the B model, it would look like the stronger condition
$K=0$ is required, which means that the canonical bundle should be trivial
even holomorphically.}
Actually, the conjecture that all \cy\ are connected
was initially formulated by Reid \cite{reid} for all {\it complex} manifolds
(and not only Calabi--Yaus) with $K=0$, extending ideas by
Hirzebruch \cite{hirzebruch}.

If now we consider a symplectic ${\cal M}$, the story is different. For one
thing, now symplectic moduli are given by $H^2({\cal M}, \Bbb R)$
\cite{moser}, so it does
not seem promising to look for a point in moduli space where three--cycles
shrink. But 2.1 in \cite{sty} shows that we can nevertheless shrink
a three--cycle symplectically, and replace it by a two--cycle. Whether
the resulting ${\cal M}'$ will also be symplectic is
not automatic, however.
This can be decided using Theorem 2.9 in \cite{sty}: the answer is yes
precisely when there is at least one relation in homology among
the three--cycles.\footnote{We should add that the relations must involve
all the three--cycles. If there is one three--cycle $A$ which is not involved
in any relation, it is possible to resolve symplectically all the other
cycles but not $A$. Examples of this type are found in \cite{vgw,werner}
when ${\cal M}$ is \ka, which is the case of interest to us and to which 
we will turn shortly. 
These examples would play in our favor, allowing us to find even more 
examples of non--\ka\  ${\cal M}'$, but
for simplicity of exposition we will mostly ignore them in the following.} 

The case of interest in this paper is actually a blending
of the two questions considered so far, whether complex or symplectic 
properties are preserved. In IIB, we will take a \cy\ ${\cal M}$ 
(which has both properties) and 
follow it in moduli space to a point at which it develops a conifold 
singularity. 
Now we perform a small resolution to obtain a manifold ${\cal M}'$ and ask 
whether this new manifold is still \ka; this question has been 
considered also by \cite{werner}. As we have seen, the complex property
is kept, and the symplectic property is not (though the question in \cite{sty}
regards more generally symplectic manifolds, disregarding the complex 
structure,
and in particular being more interesting without such a path in complex 
structure moduli space). 

Let us see why  ${\cal M}'$ cannot be \ka\ in our case. A first argument
 is not too different from an argument given
after figure \ref{fig:trans} to count four--cycles.
If the manifold ${\cal M}'$ after the transition
is \ka, there will be an element
$\omega
\in H_4$ dual to the K\"ahler form. This will have non--zero intersection
$\omega\cdot C_a= vol (C_a)$ with all the curves $C_a$ produced by the small
resolutions. Before the transition, then, in ${\cal M}$, $\omega$ will develop
a boundary,
since the $C_a$ are replaced by three--cycles $A_a$; more precisely,
$\partial \omega =\sum r^a A_a$ for some coefficients
$r^a$. This proves there will have to be at least one relation between
the collapsing three-cycles.

We can rephrase this in yet another way. 
Let us consider the case in which only 
one nontrivial three--cycle $A$ is shrinking.
Since, as remarked earlier (see figure \ref{fig:trans}),
in the compact case the curve $C$
created by the transition is trivial in homology, there exists
a three--chain $B$ such that $C=\partial B$; then we have, if $J$
is the two--form of the SU(3) structure,
\begin{equation}
0\neq \int_C J = \int_B dJ\ .
\end{equation}
Hence $dJ\neq 0$: the manifold cannot be symplectic.\footnote{ 
In the mirror picture, a similar argument shows immediately that
$d\Omega \neq 0$ on ${\cal W}'$, and hence the manifold cannot be complex.} 

Even if a symplectic $J$ fails to exist, there is
actually a non--degenerate $J$ compatible with $j$ (since 
the inclusion U(3) $\subset$ Sp$(6,\rr)$ is a homotopy equivalence, not 
unlike the way the homotopy equivalence 
O$(n)$ $\subset$ Gl$(n)$ allows one 
to find a Riemannian metric on any manifold).  
In other words, the integrable complex 
structure $j$ can be completed to a U(3) structure (and then 
to an SU(3) structure, as we will see), though not to a \ka\ one. 

This is also a good point to make some remarks about the nature of the curve
 $C$ that we will need later on. The concept of holomorphic curve 
makes sense even without an integrable 
complex structure; the definition is still 
that $(\delta +i j)^m{}_n
\partial X^n=0$, where $X$ is the embedding $C$ in ${\cal M}$. For $j$ integrable
this is the usual condition that the curve be holomorphic. But this condition
makes sense even for an almost complex structure, a fact which is expressed
by calling the curve {\it pseudo}--holomorphic \cite{gromov}. We will often 
drop this
prefix in the following. In many of the usual manipulations involving calibrated
cycles, one never uses integrability properties for the almost complex or
symplectic structures on ${\cal M}$. For example, it is still true that
the restriction of $J$ to $C$ is its volume form $\mathrm{vol}_C$. 
Exactly in the same way, one can speak of Special Lagrangian submanifolds 
even without integrability (after having defined an SU(3) structure, which 
we will in the next section), and sometimes we will qualify them as ``pseudo" 
to signify this.

Let us now consider $2\to3$ transitions. It will turn out that the results
are just mirror of the ones we gave for $3\to2$, but in this case it is
probably helpful to review them separately. After all, mirror symmetry for
complex--symplectic pairs is not as well established as for \cy s, which
is one of the motivations of the present work. (Evidence so far includes
mathematical insight \cite{kontsevich}, and, in the slightly more general
context of SU(3) structure manifolds, comparisons of four--dimensional
theories \cite{glmw,glw} and direct SYZ computation \cite{fmt}.)

Suppose now
we start (in the IIA theory) with a 
symplectic manifold ${\cal W}$ (whose moduli space
is, as we said, modeled on $H^2({\cal M}, \Bbb R)$), and that
for some value of the symplectic moduli some curves shrink. Then, it turns out
that one can always replace the resulting singularities by some
three--cycles, and still get a symplectic manifold (Theorem 2.7, \cite{sty}).
The trick is that $T^* S^3$, the deformed conifold, is
naturally symplectic, since it is a cotangent bundle.
Then \cite{sty} proves 
that this holds even globally: there is no problem in 
patching together the modifications around each conifold point. One 
should compare this with the construction used by Hirzebruch and Reid
cited above.

It is not automatic that
the resulting manifold ${\cal W}'$ is
complex, even if ${\cal W}$ is complex itself. The criterion is that
there should be at least one relation in homology between
the collapsing curves $C_a$ \cite{Fr,Ti} (see also \cite{lt} for 
an interesting application).\footnote{Actually, the criterion also 
assumes ${\cal W}$ to satisfy the $\partial \bar\partial$--lemma, 
to ensure that $H^{2,1}\subset H^3$, which is not always true on 
complex non--\ka\ manifolds; this assumption is trivially valid in the
cases we consider, where ${\cal W}$ is a \cy.}

Let us collect the transitions considered so far in a table; we also anticipate
in which string theory each transition will be relevant for us. The symmetry
among these results is clear; we will not need all of them, though.
\begin{table}[hbt]
\renewcommand{\arraystretch}{1.2}
  \begin{tabular}{c|c|c|c|c|c|}
\cline{2-6}
&transition & keeps symplectic & keeps complex & $\Delta b_2$ &$\Delta b_3$\\\hline\hline
\multicolumn{1}{|c|}{IIA} & $2\to3$ & yes (\cite{sty}, 2.7) &
if $\sum r^a_i C_a=0$ (\cite{Fr,Ti}) & $N-R$ & $2R$ \\\hline
\multicolumn{1}{|c|}{IIB} & $3\to 2$ & if $\sum r^a_i B_a =0$
(\cite{sty}, 2.9) & yes
& $R$ & $2(N-R)$\\
\hline
  \end{tabular}
\caption{The conditions for a transition to send a
complex or symplectic conifold to a complex or symplectic
manifold.}\label{table}
\end{table}

\subsection{Vacua versus geometry}
\label{sec:compare}

We can now apply the results reviewed in the previous subsection to our vacua.
Remember that in IIB we have chosen a point in moduli space in which a single
three--cycle shrinks, and in IIA one in which a single curve shrinks.

From our assumptions, the singularities affect the manifold only locally
(as opposed for example to the IIA case of \cite{ps}, in which the quantum
volume of the whole manifold is shrinking); it is hence natural to assume that
the vacua of section \ref{sec:vacua} are still geometrical.
Given the experience with the \cy\ case, it is also natural that the brane
hypermultiplet $B$ describes a surgery. But then we can use the results of
the previous subsection.

In IIB, where we have shrunk a
three--cycle, we now know that the manifold obtained by replacing the node
with a curve will be naturally complex, but will not be symplectic, since
by assumption we do not have any relations.
As we have explained, the
reason for this is that on the manifold ${\cal M}'$ after the transition,
there will be a holomorphic curve $C$ which is homologically trivial; and
by Stokes, we conclude that the manifold cannot be symplectic.

Summing up, we are proposing that in IIB the vacua we are finding are given
by a complex non--symplectic (and hence non--\ka\footnote{There might
actually be, theoretically speaking, a \ka\ structure on the manifold which
has nothing to do with the surgery. This question is natural
mathematically \cite{sty}, but irrelevant physically: such a \ka\ structure
would be in some other branch of moduli space, far from the one we are
considering, which is connected and close to the original \cy\ by
construction.}) manifold. This manifold ${\cal M}'$ is defined by a small
resolution  on the singular point of ${\cal M}$, and it has (see table
\ref{table}) Betti numbers
\begin{equation}
b_2({\cal M}')=b_2({\cal M}), ~~b_3({\cal M}')=
b_3({\cal M})-2~.
\end{equation}
In the example described in section \ref{sec:sing},
when ${\cal M}$ is the mirror of an elliptic fibration over $\Bbb F_1$,
${\cal M}'$ has $b_2=243$, $b_3/2=3$.

In IIA, a similar reasoning lets us conjecture that the new vacua correspond
to having a symplectic non--complex (and hence non--\ka) 
manifold ${\cal W}'$,
obtained from the original \cy\ ${\cal W}$
by replacing the node with a three--cycle.
This manifold ${\cal W}'$ has
\begin{equation}
b_2({\cal W}')= b_2({\cal W}) - 1,~~
b_3({\cal W}')=b_3({\cal W})~.
\end{equation}
In the example from section \ref{sec:sing},
when ${\cal W}$ is an elliptic fibration over $\Bbb F_1$, ${\cal W}'$ has
$b_2=2$, $b_3/2=244$.

Notice that these two sets of vacua are mirror by construction: we localize
in IIA and in IIB to points which are mirror to each other, and in both cases
we add the appropriate brane hypermultiplets to reveal new lines of vacua.
What is conjectural is simply the interpretation of the vacua. We now
proceed to give evidence for that conjecture.

In the IIB case, the spectrum before the transition is clearly given by
$b_3({\cal M})/2 - 1$ vector multiplets and $b_2({\cal M}) + 1$
hypermultiplets (the $``+1$" is the universal hypermultiplet).
We have seen that the potential generated by $F_3$ gives mass to one of the
vector multiplets, fixing it at a certain point in the complex moduli space.
On the other side, the number of massless hypermultiplets remains the same.
Indeed, we have added a brane hypermultiplet $B$; but this combines with the
universal hypermultiplet to give only one massless direction, the one given in
(\ref{eq:vacua}).

This is to be compared with the Betti numbers of the proposed
${\cal M}'$ from table \ref{table}: indeed, $b_2$ remains the same and $b_3$
changes by 2. Since the manifold is now non--\ka, we have to be careful in
drawing conclusions: ``\ka\ moduli'' a priori do not make sense any more, and
though complex moduli are still given by $H^{2,1}$ (by Kodaira--Spencer and
$K=0$), a priori this number is $\neq b_3/2 - 1$, since
the manifold is non--\ka.

However, two circumstances help us. The first is that, by
construction, the moduli of the manifolds we have constructed are identified
with the moduli of the singular \cy\ on which the small resolution is
performed. Then, indeed we can say that there should be
$b_3({\cal M}')/2 - 1 + b_2({\cal M}')$ complex geometrical moduli in total
(after
complexifying the
moduli from $b_2$ with periods of the anti-symmetric tensor field
appropriately, and neglecting the scalars arising from periods of RR gauge
fields).

A more insightful approach exists, and will also allow us to compare low--lying
massive states. Reduction on a general manifold of SU(3) structure (along
with a more general class which will not concern us here) has been
performed recently in \cite{glw}. (Manifolds with $SU(3)$ structures
and various differential conditions 
were also considered from the perspective of supergravity vacua, starting
with \cite{cs,gmpw}).
We have introduced a U(3) structure in the previous section as
the presence on the manifold of both a complex and a symplectic structure
with a compatibility condition. The almost complex structure $j$ allows
us to define the bundle of $(3,0)$ forms, which is called the canonical
bundle as in the integrable case. If this bundle is
topologically trivial the structure reduces further to SU(3). 
The global section $\Omega$ of the canonical bundle can actually be
used to define the almost complex structure by
\begin{equation}
T^*_{\mathrm {hol}}= \{ v_1 \in T^* | v_1 \wedge \bar\Omega=0 \}\ .
\end{equation}
The integrability of the almost complex structure is then defined by
$(d\Omega)_{2,2}=0$, something we will not always require.

Let us now review the construction in \cite{glw} from our perspective. 
In general the results of \cite{glw} require one to know the spectrum of
the Laplacian on the manifold, which is not always at hand; but in our
case we have hints for the spectrum, as we will see shortly.
We have seen that a U(3) structure, and hence also an SU(3) structure, defines
a metric. Let us see it again:
since $J\wedge \Omega=0$, $J$ is of type $(1,1)$, and then a metric can be
defined as usual: $g_{i\bar j}= -i J_{i\bar j}$.

We can now consider the Laplacian associated to this metric.
The suggestion in \cite{glw, glmw}
is to add some low--lying massive eigen--forms
to the cohomology. Since $[\Delta, d]=0$ and
$[\Delta, *]=0$, at a given mass level there will be eigen--forms of different
degrees. Suppose for example $\Delta \omega_2=m^2 \omega_2$ for a certain
$m$. Then
\begin{equation}
d \omega_2 \equiv m \beta_3
\end{equation}
will also satisfy $\Delta \beta_3=m^2 \beta_3$, and similarly
for $\alpha_3\equiv *\beta_3$ and $\omega_4 \equiv *\omega_2$. (The indices
denote the degrees of the forms.) We can repeat
this trick with several mass levels, even if coincident.

After having added these massive forms to the cohomology, we can
use the resulting combined basis to expand
$\Omega= X^I \alpha_I + \beta^I F_I$ and $J= t_i \omega_i$, formally as usual
but with some of the $\alpha$'s, $\beta$'s and $\omega$'s now being massive.
Finally, these expansions for $\Omega$ and $J$ can be plugged into certain
``universal'' expressions for the \ka\ prepotential ${\cal P}^\alpha$.
Without fluxes (we will return on this point later) and with some dilaton
factor suppressed,  this looks like \cite{glw}

\begin{equation}
  \label{eq:kprep'}
{\cal P}^1 + i {\cal P}^2 = \int d(B+iJ) \wedge \Omega,
\qquad {\cal P}^3 = \int (dC_2 - C_0 dB) \wedge \Omega.
\end{equation}

Since the reader may be confused about the interpretation of the expressions 
$\int d(B+iJ) \wedge \Omega$ and $\int (dC_2 - C_0 dB) \wedge \Omega$
which appear above (given the ability 
to integrate by parts), let us pause to give
some explanation.  Our IIB solutions indeed correspond to complex manifolds,
equipped with a preferred closed 3-form which has $d\Omega = 0$.
However, the 4d fields which are given a ${\it mass}$ by the gauging
actually include deformations of the geometry which yield $d\Omega \neq 0$,
as we discussed above.  Therefore, the potential which follows from
(\ref{eq:kprep'}) is a nontrivial function on our field space.

Let us try to apply the KK construction just reviewed to the manifold
${\cal M}'$. First of all we need some information about its spectrum.
We are arguing that ${\cal M}'$ is obtained from surgery. In \cite{dish},
it is found that the spectrum of the Dirac operator changes little, in an
appropriate sense, under surgery.
If we {\it assume} that this result goes through after twisting the Dirac
operator, we can in particular consider the Dirac operator on bispinors, also
known as the signature operator, which has the same spectrum as the Laplacian.
All this suggests that for very small $B$ and $g_s$ the spectrum on ${\cal M}'$
will be very close to the one on ${\cal M}$. Hence there will be an
eigenform of the Laplacian $\omega$ with a relatively small eigenvalue $m$
(and its partners discussed above),
corresponding to the extra harmonic forms generating $H^3$ before the
surgery.
By the reasoning above, this will also give eigenforms $\alpha$, $\beta$
and $\tilde\omega$.

Expanding now $\Omega= X^1 \alpha + \Omega_0$,
$J= t^1 \omega + J_0$, $B = b^1 \omega + B_0$ and $C_2= c^1 \omega + C_{20}$
 (where $\Omega_0$, $J_0$, $B_0$ and $C_{20}$ represent the part of
the expansion in cohomology) and using the relation $\int_{{\cal M}'} \beta_3 \wedge
\alpha_3 = 1$, we get from (\ref{eq:kprep'}):

\begin{equation}
\label{eq:kp}
{\cal P}^1 + i {\cal P}^2 \sim m (b^1 + i t^1) X^1 , \qquad {\cal P}^3
\sim m (c^1-C_0 b^1) X^1 \ .
\end{equation}

The parameter $m$ measures the non-K\"ahlerness away from the
Calabi-Yau manifold $\cal{M}$, and should be proportional to the vev of
the brane hypermultiplet $\tilde{B}_0$ of \S2.2. 
Clearly the formula is reminiscent of
the quadratic dependence on the $B$ hypermultiplet in (\ref{eq:kprep}). The 
size of the curve $C$ is measured by $t^1$.  Of course
$\tilde{B}_0$ is really a function of the $t^1$ and universal hypermultiplets.
Presumably, it and 
the massive hyper $\tilde{B}_m$ in section \ref{sec:gauged} are different
linear combinations of the curve volume and $g_s$. 
It is even tempting to map the ${\cal M}$ and ${\cal M}'$ variables 
by mapping $B$ directly to $\int_C J = t^1$, and (very reasonably) mapping 
the dilaton hypermultiplet on ${\cal M}$ directly into the one for
${\cal M}'$. Indeed, the size of $C$ would then be proportional to 
$g_s$ (at least when both are small), which is consistent 
with both being zero at the transition point. 

 Fixing this would 
require more detailed knowledge of the map between variables.
However, since the formula for the Killing prepotentials has the
universal hypermultiplet in it (which can be seen from
(\ref{eq:kp}), where $C_0$ is mixed with other hypers and some
dilaton factor is omitted in the front), it could have $\alpha'$
corrections. Moreover, (\ref{eq:kprep'}) is only valid in the
supergravity regime where all the cycles are large compared with
the string length. Hence an exact matching between the Killing
prepotentials is lacking.

We can now attempt the following comparison between the spectrum of the vacua
and the KK spectrum on the conjectural ${\cal M}'$:
\begin{itemize}
\item On ${\cal M}$, one of the vectors, $X^1$, is given a mass by the
gauging $\int F_3 \wedge \Omega$.
On ${\cal M}'$, this vector becomes a deformation of
$\Omega$ which makes it not closed, $\Omega\to \Omega+ \alpha$,
$\Delta \alpha= m^2 \alpha$. In both pictures, the vacuum is at the point
$X^1=0$. On ${\cal M}$, this is because we have fixed the complex modulus
at the point in which $A^1$ shrinks. On ${\cal M}'$, the manifold which is
natural to propose from table \ref{table} is complex, and hence $d\Omega=0$.
\item The remaining vectors are untouched by either gauging and remain
massless.
\item Both for ${\cal M}$ and for ${\cal M}'$, there are $b_2 + 1$ massless
hypermultiplets.
\item From the perspective of the gauged supergravity analysis on
${\cal M}$ there is a massive hypermultiplet too: $B$ and the
universal hypermultiplet have mixed to give a massless direction, but
another combination will be massive. On ${\cal M}'$, there is also a massive
hypermultiplet: it is some combination of $g_s$ and $t^1$, 
which multiplies the massive form $\omega$ (with 
$\Delta \omega= m^2 \omega$) in the expansion of $J$.  
To determine the precise combination one
needs better knowledge of $m(t^1,g_s)$ in (\ref{eq:kp}). 
\end{itemize}

Again, this comparison uses the fact that there is a positive eigenvalue
of the Laplacian which is much smaller than the rest of the KK tower, and this
fact is inspired by the work in \cite{dish}.

This comparison cannot be made too precise for a number or reasons. One is,
as we have already noticed, that it is hard to control the spectrum, and we
had to inspire ourselves from work which seemed relevant.
Another is that the KK reduction of ten-dimensional
supergravity on the manifold ${\cal M}'$
will not capture the full effective field
theory
precisely, as we are close (at small $B$ vevs) to a
point where a geometric transition has
occurred.  Hence, curvatures are large in localized parts of ${\cal M}'$,
though the bulk of the space can be large and weakly curved.
 And indeed, we know that
ten--dimensional type II supergravities do not allow
${\cal N}=2$ Minkowski vacua from non-K\"ahler
compactification manifolds in a regime where
all cycles are large enough
to trust supergravity
(though inclusion of further
ingredients like orientifolds, which are present in string theory,
can yield large radius ${\cal N}=2$ Minkowski vacua in this context
\cite{Orient}).
The vacua of \cite{ps}, and our own models, presumably evade this no-go
theorem via stringy corrections arising in the region localized around
the small resolution.  Some of these corrections are captured
by the local field theory analysis reviewed in \S2.3, which gives
us a reasonable knowledge of the hyper moduli space close to the
singularity.
It should be noted that the family of vacua we have
found cannot simply disappear as one increases the expectation values of
the $B$ fields and $e^{\phi}$: the moduli space of $\nn=2$ vacua is expected
to be analytic even for the fully--fledged string theory. However, new
terms in the  expansion of the ${\cal P}^\alpha$'s in terms of the $B$
hypermultiplet will deform the line; and large $g_s$
will make the perturbative
type II description unreliable.

An issue that deserves separate treatment is the following. Why have we
assumed $F_3=0$ in (\ref{eq:kprep'})? It would seem that the integral
$\int_B F_3$ cannot simply go away. Usually, in conifold transitions
(especially noncompact ones) a flux becomes a brane, as the cycle becomes 
contractible and surrounds a locus on which, by Gauss' law, there must be
a brane. This would be the case if, in figure \ref{fig:trans}, the
flux were on $A$: this would really mean a brane on $C$. In our case,
the flux is on $B$, on a chain which surrounds nothing. Without sources, 
and without being non--trivial in cohomology, $F_3$ has no choice but
disappear on ${\cal M}'$. 

\bigskip

To summarize this section, we have conjectured to which manifolds the vacua
found in section \ref{sec:vacua} correspond. In this way, we have also
provided explicit symplectic--complex non--\ka\ mirror pairs.

\section{The big picture: a space of geometries}
\label{sec:spec}

There are a few remarks that can be made about the type of complex
and symplectic manifolds that we have just analyzed, and that suggest
a more general picture. This is a speculative section, and it should be
taken as such.

One of the questions which motivated us is the following. The KK reduction
in \cite{glw} says that $\int dJ \wedge \Omega$ encodes the gauging of
the four--dimensional effective supergravity on ${\cal M}'$. Hence in some
appropriate sense (to be discussed below), $dJ$ must be integral
-- one would like
$\int dJ \wedge \Omega$ to be expressed in terms of integral combinations
of periods of $\Omega$.  This is just because the allowed gauge charges
in the full string theory form an integral lattice.
But from existing discussions, the integral nature of $dJ$ is far
from evident.
Though one can normalize the massive forms
appropriately in such a way that the expression does give an integer, this
does not distinguish between several possible values for the gauging:
it is just a renormalization, not a quantization.

Without really answering this question, we want to suggest that there must be
a natural modification of cohomology that somehow encodes some of the
massive eigenvalues of the Laplacian, and that has integrality
built in. It will be helpful to refer again to figure \ref{fig:trans}: on
${\cal M}'$ (the manifold on the right in the lower line of figure 1),
we have depicted a few relevant
chains, obviously in a low--dimensional analogy. What used to be called
the $A$ cycle is now still a cycle, but trivial in homology, as it is bounded
by a four--cycle $D$. The dual $B$ cycle, from other side, now is no longer
a cycle at all, but merely a chain, its boundary being the curve $C$.
This curve has already played a crucial role in showing that ${\cal M}'$
cannot be symplectic.

We want to suggest that a special role is played by relative cohomology
groups $H_3({\cal M}',C)$ and $H_4({\cal M}',A)$.
Remember that relative homology is
the hypercohomology of $C_\bullet(C)
\buildrel {\iota_C}\over\longrightarrow C_\bullet({\cal M}')$,
with $C_k$ being chains and the map $\iota_C$ being the inclusion. In
plain English, chains in $C_k({\cal M}',C)$ are pairs of chains
$(c_k,\tilde c_{k-1})\in C_k({\cal M}') \times C_{k-1}(C)$,
and homology is given by considering the differential
\begin{equation}
\partial(c_k, \tilde c_{k-1}) = (\partial c_k + \iota_C(\tilde c_{k-1}),
- \partial \tilde c_{k-1})\ .
\end{equation}
So cycles in $H_k({\cal M}',C)$, for example,
are ordinary chains which have boundary on $C$. $B$ is precisely
such a chain. A long exact sequence can be used to show that, when $C$ is
a curve trivial in $H_2({\cal M}')$ as is our case,
$\mathrm{dim}(H_3({\cal M}',C))= \mathrm{dim}(H_3({\cal M}'))+ 1$.
So $(B,C)$ and the usual cycles generate $H_3({\cal M}',C)$.  Similarly,
$\mathrm{dim}(H_4({\cal M}',A))= \mathrm{dim}(H_4({\cal M}'))+ 1$, and
the new generator is $(D,A)$.

Similar and dual statements are valid in cohomology. This is defined similarly
as for homology: pairs
$(\omega_k,\tilde\omega_{k-1}) \in \Omega^k({\cal M}')\times
\Omega^{k-1}(C)$, with a differential
\begin{equation}
d(\omega_k, \tilde\omega_{k-1})=(d\omega_k, \iota_C^*(\omega_k) -
d \tilde\omega_{k-1})
\end{equation}
A non--trivial element of $H^3({\cal M}',C)$ is $(0,\mathrm{vol}_C)$.
Since $C$ is a holomorphic curve, $\mathrm{vol}_C=J_{|C}\equiv
\iota_C^* J$ and hence this
representative is also equivalent to $(dJ,0)$, using the differential
above.

When we deform ${\cal M}'$ with the scalar in the massive vector multiplet
$X^1$, the manifold becomes non--complex, as we have shown in the previous
section; but
one does not require the almost complex structure to be
integrable to define an
appropriate notion of holomorphic curve.
In fact, one might expect then that, when
$d\Omega\neq0$, which corresponds to ${\cal M}'$ being non--complex,
one can also choose $A$ to be SLag (as we remarked earlier, the definition will not
really require that the almost symplectic structure be closed).\footnote{
The reader should not confuse this potential SLag, which may
exist off-shell in the IIB theory, with the pseudo-SLag 
manifold that exists on ${\cal W}'$ where $d\Omega \neq 0$ even on the
${\cal N}=2$ supersymmetric solutions.}
Definitely, the logic would hold the other way around -- if such a SLag $A$ can
be found, $\int_A \Omega\neq 0$ and then, again by integration by parts,
it follows that $d\Omega\neq 0$.

In our example, we expect the number of units $n_1$ of $F_3$ flux present 
before the transition in the IIB picture, to map to
``$n_1$ units of $dJ$'' on ${\cal M}'$.
The phrase in quotes has not been precisely defined, but
it is reasonable to think that it is defined by some kind of intersection
theory in relative homology. We will now try to make this more precise.

As we have seen, the dimension of
the relative $H_3$ can be
odd (and it is in our case), so we should not expect a pairing between
$A$ and $B$ cycles within the same group. One might try nevertheless to
define a pairing between chains in $H_3({\cal M}', C)$ and
$H_4({\cal M}', A)$; it would be defined by
\begin{equation}
(B,C)\cdot(D,A) \equiv \#(B\cap A)=\#(C\cap D)\ .
\end{equation}
In fact, if we think of another lower--dimensional analogy, in which both
$A$ and $C$ are {\it one}--dimensional in a three--dimensional manifold, it
is easy to see that what we
have just defined is a linking number between $C$ and $A$. Indeed,
$\mathrm{dim}(C)+\mathrm{dim}(A)= \mathrm{dim}({\cal M}')-1$.

This can also be rephrased in relative cohomology. Consider a bump--form 
$\delta_A$ which is concentrated around $A$ and has only components transverse 
to it, and similarly for $C$. These can be defined more precisely using 
tubular neighborhoods and the Thom isomorphism \cite{Bott}. 
Since $A$ and $C$ are trivial
in homology, we cannot quite say that these bump forms
are the Poincar\'e duals of $A$ and $C$.
But we can say that $(\delta_A,0) \in H^3(M,C)$ is the Poincar\'e dual to the
cycle $(D,A) \in H_4(M,A)$, with natural definitions for the pairing between 
homology and cohomology. $\delta_A$ is non--trivial in relative cohomology but
trivial in the ordinary cohomology $H^3(M)$, and hence there exists an 
$F_A$ such that $d F_A = \delta_A$. Then we have 
\begin{equation}
\int_{{\cal M}'} F_A \wedge \delta_C = \int_C F_A = \int_B dF_A = \#(C\cap D)
\equiv L(A,C)
\ .
\end{equation}
In other words, in cohomology we have $L(A,C)= \int d^{-1}(\delta_A) 
\wedge \delta_C$. 

Suppose we have now another form $\tilde \delta_A$ 
which can represent the Poincar\'e dual (in relative cohomology) 
to $(D,A)$. Then we can use this other form as well to compute the linking, with
identical result. This is because $(\delta_A,0)\sim (\tilde\delta_A,0)$ in 
$H^3({\cal M}',A)$ means that, by the definition of the differential above, 
$\delta_A - \tilde\delta_A= d \omega_2$ with $\omega_2$ satisfying 
$\iota_C^*\omega_2= d\tilde\omega_1$ for some form $\tilde\omega_1$ on $C$. 
Then 
\begin{equation}
\int_{{\cal M}'} d^{-1}(\delta_A - \tilde\delta_A) \wedge \delta_C= 
\int_{{\cal M}'} \omega_2 \wedge\delta_C = \int_C \omega_2 = 
\int_C d\tilde\omega_1 =0
\end{equation}
so $L(A,C)$ does not depend on the choice of the Poincar\'e dual. But now, 
remember that $(dJ,0)$ is also a non--trivial element of $H^3({\cal M}', C)$; 
if we normalize the volume of $C$ to 1, it then has an equally valid claim 
to be called a Poincar\'e dual to $(D,A)$. Indeed, $\int_{(B,C)}(dJ,0)
\equiv \int_B dJ= \int_C J= 1= 
(D,A)\cdot(B,C)$, and for all other cycles the result is 
zero. Similar reasonings apply to $d\Omega$. Then we can apply the steps above
and conclude that
\begin{equation}
L(A,C)= \int_{{\cal M}'} dJ\wedge\Omega\ .
\end{equation}
In doing this we have normalized the volumes of $C$ and $A$ to one; if
we reinstall those volumes, we get precisely that $\int dJ \wedge \Omega$ is 
a linear function of the vectors and hypers with an integral slope.

Another point which seems to be suggesting itself is the relation between
homologically trivial
Special Lagrangians and holomorphic curves on one side, and massive terms
in the expansion of $\Omega$ and $J$ on the other.
The presence of
a holomorphic but trivial curve, as we have already recalled, implies
that $dJ\neq0$: in the previous section we have seen that one actually
expects that such curves are in one--to--one correspondence with massive
eigenforms of the Laplacian present in the expansion of 
$J$ (whose coefficients represent massive fields, which vanish in
vacuum). We have argued for this
relation close
to the transition point, and for the ${\cal M}'$ that we have constructed,
but it might be that this link persists in general.
This would mean that inside an arbitrary SU(3) structure manifold,
one would have massive fields which are naturally singled out, associated to
homologically trivial holomorphic curves.

Similarly, in the IIA on ${\cal W}'$, there is a 3-cycle which is (pseudo)
Special Lagrangian but homologically trivial. 
Its presence implies that $d\Omega \neq 0$, in keeping with the fact
that the IIA vacua are non-complex.   

Reid's fantasy \cite{reid}\ involved the conjecture that by shrinking
-1 curves, and then deforming, one may find a connected configuration
space of complex threefolds with $K=0$.
Here, we see that it is natural to extend this fantasy to include
a mirror conjecture: that the space of symplectic non-complex manifolds
with SU(3) structure is similarly connected, perhaps via transitions
involving the contraction of (pseudo) Special Lagrangian cycles, followed
by small resolutions.  The specialization to -1 curves in \cite{reid}\
is probably mirror to the requirement that the SLags be rigid, in
the sense that $b^1=0$.

In either IIB or IIA, we have seen that (at least close to the
transition) there is a natural
set of massive fields to include in the low-energy theory, associated
with the classes of cycles described above. 
Allowing these fields to take on expectation values may allow one to
move off-shell, filling out a
finite--dimensional (but large) configuration space, inside
which complex and symplectic
manifolds would be zeros of a stringy effective potential.
While finding such an ${\cal N}=2$ configuration space together with
an appropriate
potential to reveal all ${\cal N}=2$ vacua is clearly an ambitious goal,
it may also provide a fruitful warm-up problem for
the more general question of
characterizing the string theory ``landscape'' of ${\cal N} \leq 1$ vacua
\cite{landscape}.

In this bigger picture, this paper is a Taylor expansion of the 
master potential around a corner in which the moduli space of ${\cal M}'$
meets the moduli space of compactifications on
${\cal M}$ with RR flux.


{\bf Acknowledgments:} We would like to thank P.~Aspinwall, B.~Florea
and A.~Kashani-Poor for useful discussions, and I.~Smith and R.~Thomas for
some patient explanations of their work.
The authors received support from the DOE under contract
DE-AC03-76SF00515 and from the National Science Foundation under grant
0244728.
SK was also supported by a David and Lucile Packard Foundation Fellowship
for Science and Engineering.

\appendix
\section{Details about an example}
\label{app:toric}
We will detail here the transition for the example mentioned in 
section \ref{sec:sing}. We will do so on the IIA side, which is 
the one which involves the strictest assumptions, as explained there. 

The \cy\ ${\cal W}$ is an elliptic fibration over the Hirzebruch surface
$\Bbb F_1$. It is convenient to describe it as a hypersurface in a toric
manifold $V$. The fan for the latter is described by the columns of
the matrix
\begin{eqnarray*}  
&\begin{array}{ccccccc}
 v_1 & v_2 & v_3 & v_4 & v_5 & v_6 & v_7 \\
\end{array}&\\
&\left[\begin{array}{ccccccc}
 0 & 0 & 0 & 1 & 0 &-1 & 0\\
 0 & 0 & 1 & 1 &-1 & 0 & 0\\
 0 &-1 & 2 & 2 & 2 & 2 & 2\\
-1 & 0 & 3 & 3 & 3 & 3 & 3
\end{array}
\right]&\ .\\
\end{eqnarray*}
The last five vectors lie in the same plane, determined by
the last two coordinates; let us plot the first two coordinates, along 
with three different triangulations:

\begin{picture}(200,80)(-20,-10)
\put(30,0){\line(0,1){30}}
\put(30,0){\line(-1,1){30}}
\put(30,0){\line(1,2){30}}
\put(30,30){\line(-1,0){30}}
\put(30,30){\line(1,1){30}}
\put(30,60){\line(-1,-1){30}}
\put(30,60){\line(1,0){30}}
\put(30,30){\line(0,1){30}}
\put(80,30){\vector(1,0){30}}
\put(180,0){\line(0,1){30}}
\put(180,0){\line(-1,1){30}}
\put(180,0){\line(1,2){30}}
\put(180,30){\line(-1,0){30}}
\put(180,30){\line(1,1){30}}
\put(180,60){\line(-1,-1){30}}
\put(180,60){\line(1,0){30}}
\put(170,35){$v_7$}
\put(140,25){$v_6$}
\put(180,-10){$v_5$}
\put(175,65){$v_3$}
\put(215,65){$v_4$}
\put(230,30){\vector(1,0){30}}
\put(330,0){\line(0,1){30}}
\put(330,0){\line(-1,1){30}}
\put(330,0){\line(1,2){30}}
\put(330,30){\line(-1,0){30}}
\put(330,30){\line(1,1){30}}
\put(330,60){\line(-1,-1){30}}
\put(330,60){\line(1,0){30}}
\put(300,30){\line(2,1){60}}
\end{picture}

The vectors of the fan are indeed the right ones to describe the 
$\Bbb F_1$ base. The fan is further specified by the higher--dimensional
cones in the picture, with the first triangulation really describing 
the elliptic fibration over $\Bbb F_1$, the last describing
a space related to the first by 
a flop, and the middle triangulation describing the singular case. (The points
have been labeled in the singular case only to avoid cluttering the picture.)
We associate as usual a homogeneous coordinate $z_i$ to each of the $v_i$'s,
with charge matrix given by the (transposed) kernel of the matrix above:
\[
\left[
\begin{array}{ccccccc}
0 & 0 & 3 &-2 & 1 &-2 & 0\\
6 & 4 & 1 & 0 & 1 & 0 & 0\\
3 & 2 & 0 & 0 & 0 & 0 & 1
\end{array}\right]
\]

From the picture we see that the flopped locus in $V$ lies at 
$z_3=z_4=z_6=z_7=0$. One has to check whether this locus intersects the
\cy\ only once. This is done by looking at the equation for ${\cal W}\subset
V$, which for a certain point in the complex moduli space reads 
$z_1^2 + z_2^3 + z_3^{12} z_4^{18}z_7^6 + z_5^{12} z_6^6 z_7^6 +
z_3^{12} z_6^{18} z_7^6 + z_4^6 z_5^{12} z_7^6=0$; hence
we get the singular locus $z_1^2 + z_2^3=0$ on ${\cal W}$. Taking into account
the $\cc^*$ actions,
this corresponds to only one point $p$ as desired. 
To verify that the normal bundle of the shrinking curve has charges $(-1,-1)$, 
one can identify the combination of the charges that keeps $p$
invariant; this action turns out to be $(1,1,\lambda, \lambda^{-1},1,
\lambda^{-1}, \lambda)$, $\lambda\in \cc^*$, 
which is the right one for a conifold point.

\end{document}